\documentclass[10pt,conference,letterpaper]{IEEEtran}
\IEEEoverridecommandlockouts

\usepackage{times}
\usepackage[]{todonotes}

\usepackage{cite}
\usepackage{amsmath,amssymb,amsfonts}
\usepackage{algorithmic}
\usepackage{graphicx}
\usepackage{textcomp}
\usepackage{xcolor}
\usepackage{alltt} 

\def\BibTeX{{\rm B\kern-.05em{\sc i\kern-.025em b}\kern-.08em
    T\kern-.1667em\lower.7ex\hbox{E}\kern-.125emX}}

\usepackage{array}
\usepackage{mdwtab}
\usepackage{hyperref}
\usepackage{booktabs}
\usepackage{flushend}

\usepackage{listings}
\usepackage{color}

\definecolor{dkgreen}{rgb}{0,0.6,0}
\definecolor{gray}{rgb}{0.5,0.5,0.5}
\definecolor{mauve}{rgb}{0.58,0,0.82}

\begin{document}
\pagenumbering{arabic} 
\pagestyle{plain}



\title{Implementation of Smart Contracts Using Hybrid Architectures with On- and Off-Blockchain Components \\ 
\vspace{0.3cm}
\normalsize{(Extended Version: 31 Jul 2018)}}

\author{
\IEEEauthorblockN{Carlos Molina-Jimenez}
\IEEEauthorblockA{Computer Laboratory\\
University of Cambridge, UK\\
carlos.molina@cl.cam.ac.uk}\\   

\and

\IEEEauthorblockN{Ioannis Sfyrakis}
\IEEEauthorblockA{School of Computing\\
Newcastle University, UK\\
ioannis.sfyrakis@ncl.ac.uk}\\

\and

\IEEEauthorblockN{Ellis Solaiman\\}
\IEEEauthorblockA{School of Computing\\
Newcastle University, UK\\
ellis.solaiman@ncl.ac.uk}

\and

\IEEEauthorblockN{Irene Ng\\}
\IEEEauthorblockA{Hat Community Foundation\\
Cambridge, UK\\
irene.ng@hatcommunity.org}

\and

\IEEEauthorblockN{Meng Weng Wong}
\IEEEauthorblockA{CodeX, Stanford University\\
mengwong@stanford.edu\\
Legalese.com \\
mengwong@legalese.com}

\and

\IEEEauthorblockN{Alexis Chun\\}
\IEEEauthorblockA{Visiting Fellow, \\
Singapore Management University \\
Legalese.com\\
alexis@legalese.com
}

\and

\IEEEauthorblockN{Jon Crowcroft\\}
\IEEEauthorblockA{Computer Laboratory\\
University of Cambridge, UK\\
jon.crowcroft@cl.cam.ac.uk\\}
}

\maketitle
\thispagestyle{empty}

\begin{abstract}

Recently, decentralised (on-blockchain) platforms have emerged to complement centralised (off-blockchain) platforms for the implementation of automated, digital (``smart'') contracts. However, neither alternative can individually satisfy the requirements of a large class of applications. On-blockchain platforms suffer from scalability, performance, transaction costs and other limitations. Off-blockchain platforms are afflicted by drawbacks due to
their dependence on single trusted third parties. We argue that in several application areas, hybrid platforms composed from the integration of 
on- and off-blockchain platforms are more able to support smart contracts that deliver the desired quality of service (QoS). Hybrid architectures are largely unexplored. To help cover the gap, in this paper we discuss the implementation of smart contracts on hybrid architectures. As a proof of concept, we show how a smart contract can be split and executed partially on an off-blockchain contract compliance checker and partially on the Rinkeby Ethereum network. To test the solution, we expose it to sequences of contractual operations generated mechanically by a contract validator tool.

\end{abstract}


\section{Introduction}
\noindent This paper investigates scenarios involving two or more commercial parties interacting digitally with each other in a relationship regulated by some computer-readable formal specification that details the operational aspects of the parties' business with each other. If this specification were written in natural language and signed on paper by the parties, it would be considered a traditional legal contract enforceable by a court. However, a specification written in a formal language, intended for digital execution and performance by the parties, constitutes a new breed of contract. The fact that such contracts may be executed, performed, and enforced by technology alone promises to largely obviate the need for ``offline'' court enforcement. Hence the appellation `smart' contract.

Approaches to automated contract execution pre-date today's on-blockchain Bitcoin and Ethereum contracts. For decades, financial institutions have executed trades digitally; Nick Szabo~\cite{Szabo1997} used the term "smart contract" prior to Satoshi~\cite{Satoshi2008}; and purely mechanical vending machines have sold cold drinks and train tickets long enough for Lord Denning~\cite{Denning1970} to remark in 1970:

\begin{quote}
The customer pays his money and gets a ticket. He cannot refuse it. He cannot get his money back. He may protest to the machine, even swear at it. But it will remain unmoved. He is committed beyond recall. He was committed at the very moment when he put his money into the machine. The contract was concluded at that time.
\end{quote}

\noindent This quote is as relevant to the smart contracts of today as it was to the train tickets of 1970, with one key difference---in 1970 the contract would still have to be \emph{performed} in the real world (the customer gets on the train, which moves him to his destination), whereas today a smart contract could be performed entirely digitally, without human involvement.

This paper uses the running example of an online data sale, which translates the operational essence of a traditional sale and purchase agreement toward smart-contract digital execution. A seller offers some digital content; a buyer pays; the seller delivers. The traditional, natural language version of such a contract might say \emph{The data seller is obliged to make the purchased data available for retrieval by the buyer for a term of 3 days after payment is received}. Such clauses are clear candidates for formalisation and digital execution by means of smart contracts.

A \textbf{smart contract} is an executable program (written in some programming language like, Java, C++, Solidity, Go, etc.) that is deployed to mediate contractual interactions between two or more parties. Its task is to detect (and if possible prevent) deviations from the agreed upon behaviour. To perform its task, the smart contract i) intercepts each business event generated by the parties, ii) analyses it to determine if it is contract compliant or not, iii) produces a verdict, and iv) records the outcome in an indelible log that is available for verification, for example, to sort out disputes. Notice that in some
applications, the declaration of the verdict is directly and intricately associated to an action (for example, collect the payment) that is executed when the verdict is positive. In this paper we separate the two acts and focus on the verdict. We argue that 
the verdict is the most crucial task and the essence of smart contracts. The subsequent action is an arbitrary reaction to the verdict and can be immediately or eventually executed by the smart contract or by another component of the application. 

We regard a smart contract as a piece of middleware
expected to deliver a service with some QoS. Examples of QoS are: trust (who can be trusted with the deployment of the smart contract), transparency (can the contracting and third parties verify the verdicts), throughput (the number of operation that the smart contract can verify per second), response time (the time it takes to output a verdict), transaction fees (the monetary cost that the parties pay to the smart contract for processing each operation). Different applications (for example, a buyer--seller contract, property renting contract, etc.) will demand different QoS. The question here: what technology to use to implement smart contracts that satisfy the imposed requirements. Note that in this paper
we use the terms smart contract and contract synonymously.

Centralised (off-blockchain) and decentralised (on-blockchain) 
platforms are available for the implementation of smart contracts. However, we argue that neither alternative can individually provide the QoS demanded
by some applications.

Currently, leading examples of smart-contract blockchain platforms are Bitcoin~\cite{AndreasAntonopoulos2017}, Hyperledger~\cite{HyperledgerHome} and Ethereum~\cite{Ethereum2017}. Bitcoin has been criticised for throughput limitations: it can only process 7 transactions per second, compared to Visa's 2000 transaction per second~\cite{Trent2016}. And it takes Bitcoin about 10 minutes to publish a transaction in its block~\cite{Noyen2014}.

Off-blockchain platforms were available long before
the Satoshi's seminal paper~\cite{Santosh2005} that launched Bitcoin~\cite{Minsky1985,Lindsay1993,MilosevicJosang2002, PedroGama2005,perringodart,ludwig2003soa,LaiXu2004, PedroGama2005, MolinaTSC2011}. These platforms rely on Trusted Third Parties (TTP) which may not deserve that trust.

The central argument of this paper is that in several application areas, hybrid platforms composed from the integration of off- and on-blockchain platforms are better~\cite{CarlosIoannisTurin2018} than either alone.
To date, the use of hybrid architectures in smart contract implementations has been largely unexplored. This paper aims at helping to close the research gap. 

The main contribution of this paper is
the implementation of a smart contract on a hybrid architecture.
 At this stage we aim at proving the concept rather than at evaluating performance.  We show how a smart contract can be split and executed partially on an off--blockchain contract compliance checker and partially on an Ethereum blockchain. To test the solution, we expose it to sequences of contractual operations generated mechanically by a contract validator tool.

We continue the discussion as follows: We present a contract example as a motivating
scenario in Section~\ref{motivatingscenario}. We discuss different approaches to smart contract implementation in Section~\ref{implementationalternatives}. Our experience in the implementation of the hybrid architecture is discussed in Section~\ref{implhybridarchitecture}. In Section
\ref{exceptionaloutcomes} we present an execution
model of contractual operation that accounts for
exceptional completions. In Section~\ref{relatedwork} we place our research in context. In Section~\ref{futureresearch} we 
discuss open research questions and pending work. In Section~\ref{conclusions}, we 
present concluding remarks.

\section{Motivating scenario}
\label{motivatingscenario}
\noindent Alice sells data that she has aggregated from different sources (domestic sensors, social networks, shopping, etc.) and stored in a repository, as envisioned in the HAT project~\cite{HATHome}. Bob (the ``buyer'') buys data from Alice (the ``seller'' or ``store''). The contract that governs their relationship includes the following clauses.
 
\begin{it}
\begin{enumerate}
\item The buyer (Bob) has the \textbf{right} to place a \textbf{buy request} with the store to buy an item.
\item The store (Alice) has the \textbf{obligation} to respond with either \textbf{confirmation} 
       or \textbf{rejection} within 3 days of receiving the request.
 \begin{enumerate}
   \item  No response from the store within 3 days will be treated 
          as a rejection.
 \end{enumerate}
\item The buyer has the \textbf{obligation} to either \textbf{pay} or \textbf{cancel} the  request
       within 7 days of receiving a confirmation.
      \begin{enumerate}
        \item No response from the buyer within 7 days will be 
               treated as a cancellation.
      \end{enumerate}
  \item The buyer has the \textbf{right} to \textbf{get a voucher} from the store,
        within 5 days of submitting payment.
\item If, within 3 days of receiving the voucher, the buyer presents the voucher to the store, then the store \textbf{must deliver} the requested item.
\end{enumerate}
\end{it}

\noindent The clauses include contractual operations (for example, \textbf{buy request}, \textbf{reject} and \textbf{confirmation}) that the parties have the right or obligation to execute under strict time constrains to honour the contract. We have highlighted the operations in bold. Though the clauses are relatively simple, they are realistic enough to illustrate our arguments.

\section{Implementation alternatives}
\label{implementationalternatives}
\begin{figure}[!t]
	\centering
	\includegraphics[width=0.95\columnwidth]{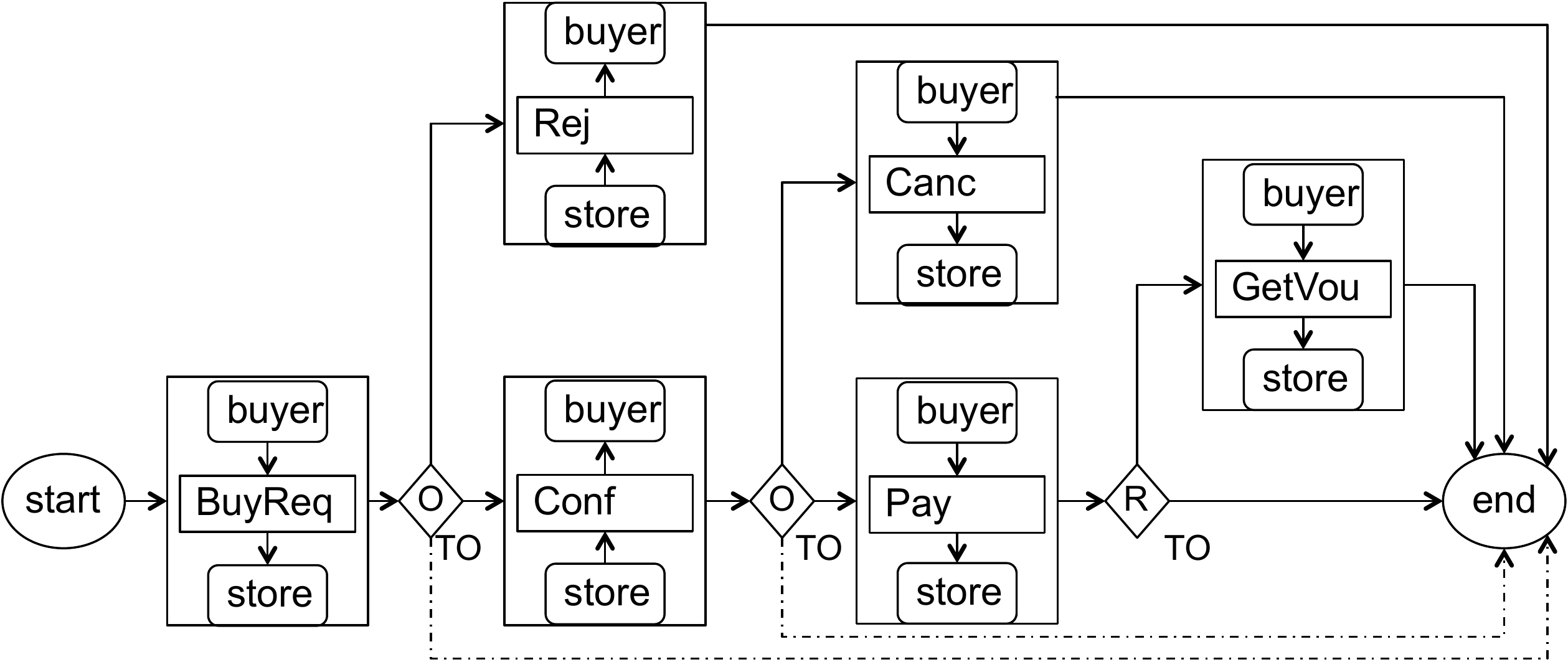}
	\caption{A contract between a buyer and store for trading personal data.}
	\label{fig:buyerstorecontractmsgs}
\end{figure}

\noindent A close examination of the example reveals that
the clauses describe the set of legal execution paths that the interaction between the two parties can follow. As such, the contract written in English---pseudocode---can be converted into a smart contract---a formalism---that can be enforced mechanically. To show how this can be done, we convert the English text contract into a systematic notation. Fig.~\ref{fig:buyerstorecontractmsgs} shows a graphical view of the contract example.

The operations in the English contract have been mapped to messages sent by one party to another. For example, the execution of the operation \textbf{buy request} corresponds to the \textbf{BuyReq} message sent by the buyer to the store. Similarly, the execution of the operation \textbf{reject} corresponds to the \textbf{Rej} message sent by the store to the buyer.
The diamonds represent exclusive splits in the execution path and have been labeled with \emph{O} (Obligation) and \emph{R} (Right).
\emph{TO} stands for Time Out and is used for defaults. Failure to execute and obligatory operations results in abnormal contract end (represented by dashed lines) with disputes to be sorted off line.

Fig.~\ref{fig:buyerstorecontractmsgs} reveals that the contract example can be modelled and implemented as a finite state machine (FSM). The challenge for the developer is to select a suitable architecture and technology for implementation. As discussed in~\cite{CarlosIoannisTurin2018}, there are several approaches to smart contract implementations:

 \begin{itemize}
  \item Centralised: The smart contract is deployed on a Trusted 
        Third Party. This approach is also  known as 
        off-blockchain implementation since there is no 
        blockchain involved. 
  \item Decentralised: The smart contract is deployed on a 
        blockchain platform such as Ethereum. This approach is
        also known as on-blockchain.
  \item Hybrid: The contract is split and deployed partly off- and
         partly on-blockchain. Some clauses are enforced
         off-blockchain; others are enforced on-blockchain.  
         The partition 
         is based on several criteria including blockchain cost,  
         performance, consensus latency, smart contract languages 
         and privacy. See~\cite{CarlosIoannisTurin2018}
         ~\cite{Eberhardt2017} and ~\cite{Guy2015}. 
 \end{itemize}

\section{Hybrid architecture}
\label{implhybridarchitecture}
\noindent As explained in~\cite{CarlosIoannisTurin2018}, the alternatives
discussed in Section~\ref{implementationalternatives} offer different 
QoS attributes (for example, scalability, privacy, consensus latency, transaction fees) that render them suitable for some applications but unsuitable for others. There exist applications whose requirements are only met by the hybrid
approach. In this section, we demonstrate a hybrid implementation.

Complexity
 inevitably emerges from the interaction between the off-blockchain and 
 on-blockchain components. Several architectures are possible, such as master--slave or peer--to--peer. Alternatively, we can place them in a
 parallel-pipe relationship where an off-blockchain smart contract is deployed by one
 of the contractual parties to mirror the work of the on-blockchain 
 smart contract, say to double-check its outputs. Other deployment alternatives are discussed in~\cite{MolinaSOCA2011}. As discussed below, interaction intricacies demand
 systematic scrutiny to prevent buggy smart contracts.
 
  
  The central idea of the hybrid approach is to 
  divide the contractual operations into off-blockchain operations 
  and on-blockchain operations. Off-blockchain operations are evaluated for contract compliance by a centralised smart contract deployed on a trusted third party. In contrast, on-blockchain operations are evaluated by a decentralised smart contract deployed on a blockchain.
  
\begin{figure}[!t]
	\centering
	\includegraphics[width=0.85\columnwidth]{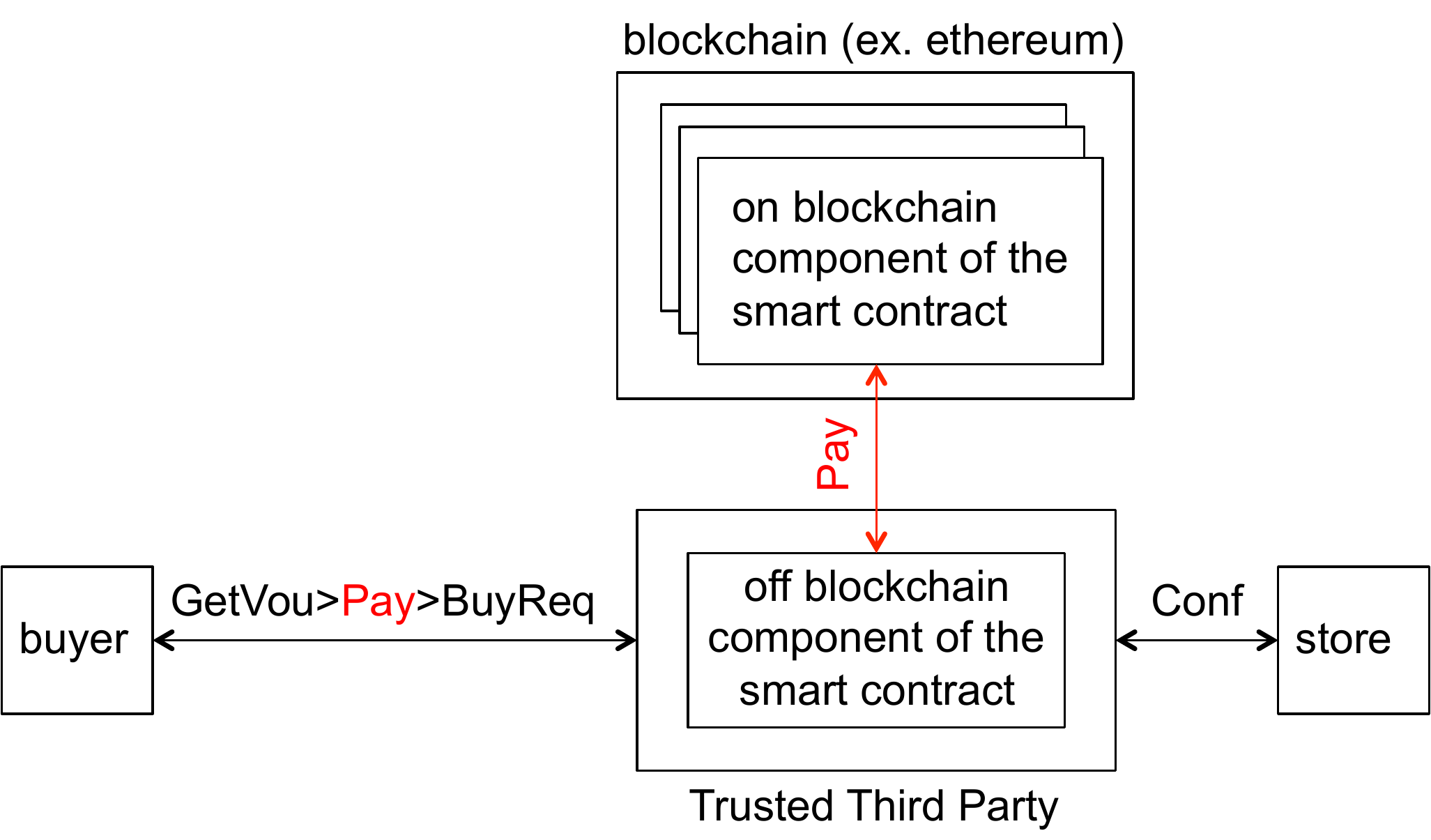}
	\caption{Smart contract split into on- and off-blockchain enforcement.}
	\label{fig:smartcontractsplit}
\end{figure}

Let us assume henceforth that Alice and Bob have agreed to use a hybrid architecture where
the operation {\color{red}\texttt{Pay}} will be enforced on-blockchain and all other operations, off-blockchain. 
An abstract view of the corresponding hybrid architecture is shown in Fig.~\ref{fig:smartcontractsplit}. 
 
\begin{figure}[!t]
	\centering
	\includegraphics[width=0.95\columnwidth]{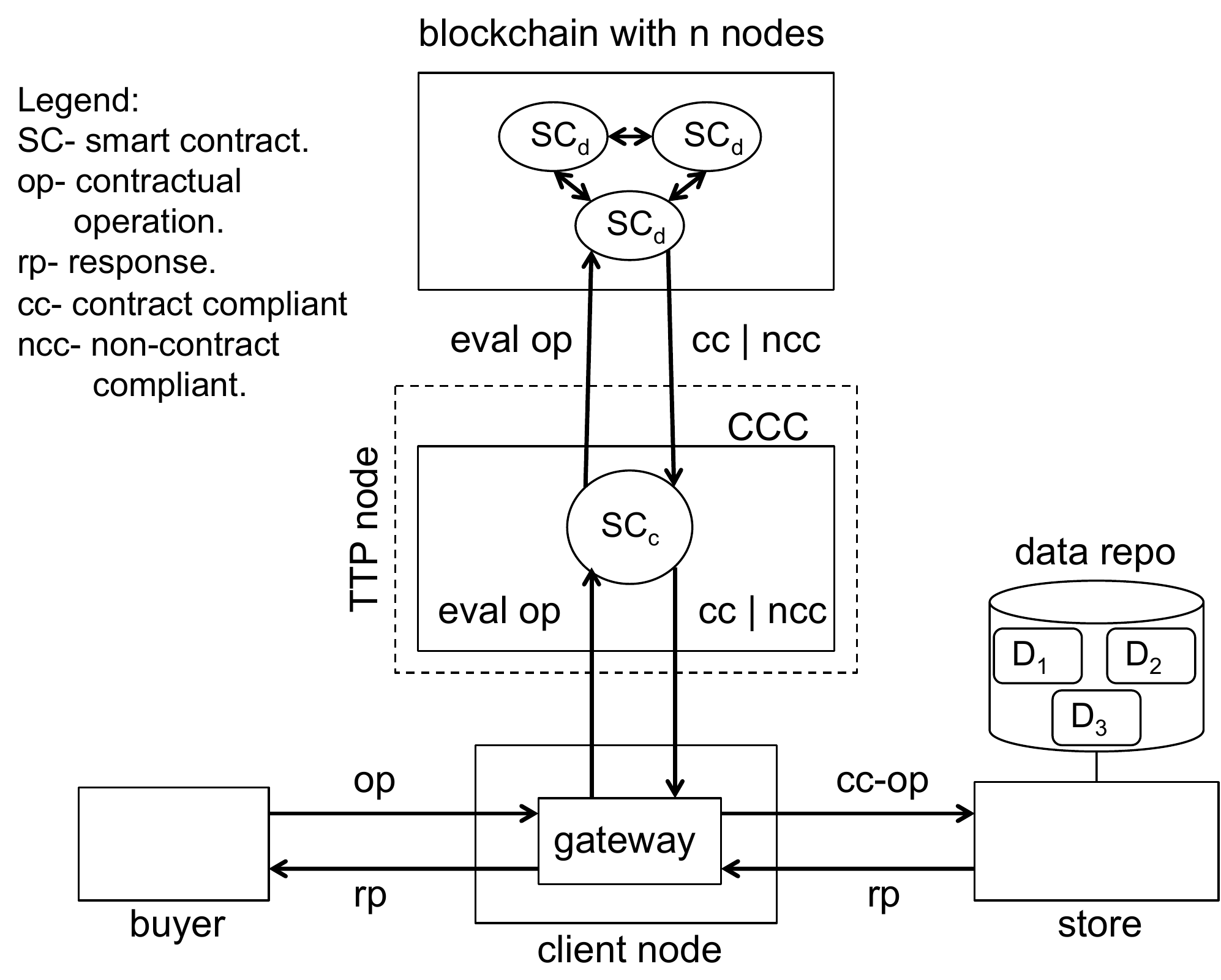}
	\caption{A hybrid architecture for smart contracts: conceptual view.}
	\label{fig:conceptualarchitecture}
\end{figure}

Fig.~\ref{fig:conceptualarchitecture} applies the concepts of Fig.~\ref{fig:smartcontractsplit} to the contract example. $D_1$,
 $D_2$ and $D_3$ are pieces of personal data that Alice is willing
 to sell, presumably under different conditions of price,
 privacy and so on.
 
The hybrid architecture can be implemented using
a range of technologies. To realise the on-blockchain, decentralised component, we use the
Rinkeby testnet of the Ethereum blockchain ~\cite{Ethereum2017}. To realise the
off-blockchain centralised component, we use the latest version 
of the Contract Compliance Checker (CCC) 
developed by University of Newcastle~\cite{conch}. The 
integration follows a master--slave
 relationship between the centralised and decentralised
 smart contract components where the former is ``in charge''. The on-blockchain smart contract reads input events from the off-blockchain contract, treating it as an oracle. The off-blockchain code is able to read on-blockchain events---the chain itself.

 \subsection{Contract compliance checker}
 \label{ccc}
\noindent We use the contract compliance checker~\cite{MolinaTSC2011,conch} 
 because it offers several features
that can ease integration with a blockchain platform. The CCC
is an open source tool designed for the enforcing of
smart contracts. It is a Java application composed
of several files, RESTful interfaces, and a database.
At its core lies a FSM that grants and removes \textbf{rights},
\textbf{obligations} and \textbf{prohibitions} from the contracting 
parties as the execution of the contract progresses. To 
enforce a smart contract with the CCC, the developer (i) writes the contract in the Drools language and stores it in a
\emph{.drl} file (for example \emph{dataseller.drl}), (ii) 
loads (copies) the  \emph{drl} file
into the \emph{configuration/drools/upload} folder,
and (iii) instantiates the CCC as a web server (for
example on a TTP node) that waits for the arrival
of events representing the contractual operation. An
\emph{event} is a notification about the execution of
a contractual operation by a contractual partner. For example
when the buyer of Fig~\ref{fig:buyerstorecontractmsgs}, 
executes the operation \emph{BuyReq} the event
\emph{BuyReq} is generated by the buyer's application and
sent to the CCC for evaluation. Similarly, when the seller executes
the operation \emph{Conf}, the seller's application sends the event \emph{Conf} to
the CCC for evaluation.
 
Drools is a declarative, Turing-complete language 
designed for writing business rules~\cite{Drools5.3.0UserGuide}.
The contract loaded to the CCC is capable of evaluating contractual 
operations issued by business partners as RESTful requests against its
rules. Rules give RESTful responses that can be the outcome 
 of an evaluation of an operation (\emph{contract compliant} or 
 \emph{non contract compliant}) or an arbitrary message such as a 
 request to execute an operation on a blockchain.

\subsection{Client node}
\noindent The client node is an ordinary node and not necessarily
the same as the TTP shown in the figure. It is
responsible for hosting the \emph{gateway}.  
Contractual operations (\emph{op}) are initiated by the business parties, such as \emph{BuyReq}, and \emph{Pay}. 
The $SC_c$ contract determines if a given operation is contract compliant (\emph{cc}) or non contract compliant (\emph{ncc}). 
The $SC_c$ is in control of the \emph{gateway}
which grants access to the seller's data. For example, when the buyer wishes to
access the seller's data: (i) the buyer issues the corresponding operation
against the gateway, (ii) the gateway forwards the operation to the $SC_c$, (iii) the $SC_c$ evaluates the operation in
accordance with its business rules that encode the contract 
clauses and responds with either \emph{cc} or \emph{ncc} to open or 
close the gateway, respectively, and (iv) the opening of the gateway allows the buyer's operation to reach the data repository and retrieve the 
response (\emph{rp}) that travels to the buyer. Note that, to keep the figure simple, the arrows show only the direction followed by operations initiated by the buyer. Operations initiated by the seller follow a similar procedure but right to left.

\subsection{Ethereum}
\noindent We chose the Ethereum platform~\cite{Ethereum2017} to implement the decentralised contract enforcer for the following
reasons: It is currently one of the most mature blockchains. It supports Solidity---a Turing--complete language~\cite{Solidity2017} that designers can use for encoding stateful smart contracts of arbitrary complexity. For complex contracts, Ethereum is more convenient 
than Bitcoin which supports only an opcode stack-based script language~\cite{BitcoinScript}. In addition, Ethereum offers developers on-line compilers of Solidity code~\cite{remix}.
Equally importantly, Ethereum provides, in addition to the main Ethereum network (Mainnet), four experimental 
networks (Ropsten, Kovan, Sokol and Rinkeby) that developers can use for experiments using
Ethereum tokens instead of ``real'' ether money~\cite{RopstenKovanRinkeby,web3jtransactions}. We run our experiments in Rinkeby as it is the most stable testnet. 
To pay for the transaction to deploy the smart contract on
Rinkeby we could have created ERC20--compatible
ether tokens~\cite{CreateERC20token}. However, we opted 
for simplicity and used ethers (tokens) requested from faucet~\cite{Faucet}.
These ethers are free and can be requested by anybody
with a wallet address and a third party social network accounts such as
Twitter and Face Book.

\subsection{Execution sequences for testing the hybrid architecture} 
\label{modelcheckingtesting}
\noindent A feature of on-blockchain contracts is that
because of their decentralisation and openness, they are generally immutable after deployment. Therefore, we suggest that smart contracts
should be thoroughly validated (for example, using conventional model checking tools) to uncover potential logical inconsistencies
in their clauses (omissions, contradictions, duplications, etc.) \cite{MolinaMWSOC2009}. In addition, we suggest that the actual implementation be systematically tested before deployment. These tasks demand the assistance of software tools, such as the \emph{contraval} tool that we have developed~\cite{contraval}, specifically for model checking and testing contracts~\cite{Abdelsadiq2010,contraval}. It is based on
the standard Promela language and the Spin model checker. It supports epromela (an extension of Promela) that provides constructs for intuitively expressing  
and manipulating contractual concepts such rights, obligations and
role players.
 
In this work, we use \emph{contraval} for model checking the contract example 
and, more importantly, for generating the execution sequences that we use
for testing the hybrid architecture of Fig.~\ref{fig:cccweb3jethereum}. 
We define an \textbf{execution sequence} as a set of one or more
contractual operations that the contractual parties need to execute
to progress the smart contract from the \emph{start} to the
\emph{end}.

Our proof-of-concept proceeded as follows:

\begin{enumerate}
 \item We converted the clauses of the contract example into a formal model written
       in epromela. We called it \emph{dataseller.pml}.
  \item We model-checked the contractual model with Spin to
        verify conventional
       correctness requirements (deadlocks, missing messages, etc.) and 
       typical contractual problems (clause duplications, 
       omissions, etc.)~\cite{MolinaMWSOC2009}).    
 \item We augmented the contractual model with an LTL
       formula and exposed it to Spin, and instructed Spin to produce counterexamples
       containing execution sequences of interest.
  \item We ran a Python parser (called \emph{parser-filtering.py}) that 
       we have implemented, over
       the Spin counterexamples to extract the execution sequences.

\end{enumerate}

A close examination of Fig.~\ref{fig:buyerstorecontractmsgs} will reveal
that it encodes six alternative paths from contract \emph{start} 
to contract \emph{end}.  

{\scriptsize
\begin{verbatim}
// Execution sequences encoded in Fig 1.
// RejConfTo=Rej or Conf timeout, 
// PayCancTo=Pay or Canc timeout
 
 seq1: {BuyReq, Rej}
 seq2: {BuyReq, Conf, Canc}
 seq3: {BuyReq, Conf, Pay}
 seq4: {BuyReq, RejConfTo}
 seq5: {BuyReq, Conf, PayCancTo}
 seq6: {BuyReq, Conf, Pay, GetVou} 
\end{verbatim}
}

In Section~\ref{determination-cc-ncc} we show how these
sequences can be analysed systematically by the smart contracts of
Fig.~\ref{fig:cccweb3jethereum}. However, a visual analysis reveals that
Seq1, Seq2 and Seq3 result in normal contract completion. However,
Seq4 and Seq5 result in abnormal contract completion. In Seq4 the
seller fails to meet its obligation (to execute either \emph{Rej} or \emph{Conf})
before the 3 day deadline. Similarly, in Seq5, the buyer fails to execute
either \emph{Pay} or \emph{Canc} before the 7 day deadline.
 Observe that although the buyer has 5 days to claim a voucher, failure to execute
 \texttt{GetVou} does not result in abnormal contract completion
 because \texttt{GetVou} is a right, rather than an obligation.
 Seq6 is particularly problematic. It will be
 analysed separately in Section~\ref{determination-cc-ncc}.

\begin{figure}[!t]
	\centering
	\includegraphics[width=0.75\columnwidth]{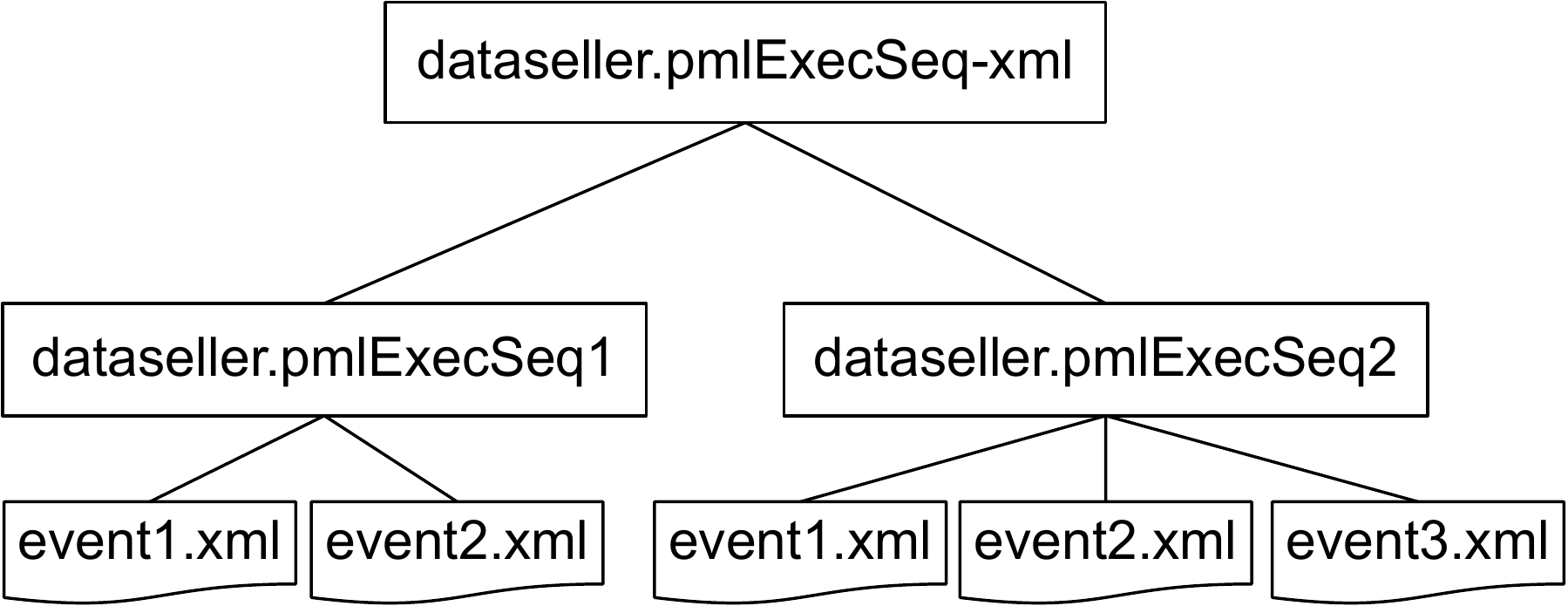}
	\caption{Folder with execution sequences of the contract example.}
	\label{fig:folderwithexecsequences}
\end{figure}

 To ease sequence manipulation, we programmed the Python
 parser to store the execution sequences in 
 $N$ subfolders, one for each sequence. In
 our experiments with the contract example, we used the
 folders shown in Fig.~\ref{fig:folderwithexecsequences}.
            
      Each subfolder  \emph{dataseller.pmlExecSeqi} (only two are shown in the 
      figure) contains $M$ files, \emph{event1.xml}, \emph{event2.xml}, etc., one
      file for each event included in the sequence.
      In our experiments, the subfolder \emph{dataseller.pmlExecSeq1} is
      related to Seq1 and consequently,
      containes two files: event1.xml and event2.xml that correspond,
      respectively, to the \emph{BuyReq} and \emph{Rej} events.
 
       We use XML-like tags to enrich the events with additional information, 
       The contents of files  event1.xml and event2.xml are shown
       in the following code, left and right, respectively.

 \parbox{1.8in}{\scriptsize{\begin{alltt}
<event>\\
<origin>buyer</origin> \\
<respond>store</respond> \\
<type>BuyReq</type> \\
<status>success</status>\\
</event>\end{alltt}}}\parbox{1.8in}{\scriptsize{\begin{alltt}
<event>\\
<origin>store</origin> \\
<respond>buyer</respond> \\
<type>Rej</type> \\
<status>success</status>\\
</event>
\end{alltt}}}

The \emph{type} tag indicates the type of the
event. For example, the execution of the contractual operation
\emph{BuyReq} produces an event of type \emph{BuyReq},
similarly, the execution of the contractual
operation \emph{Rej} produces an event of type \emph{Rej}.
The \emph{origin} tag 
indicates the party that originated  the event (the buyer in
the example of the left),
similarly, \emph{respond} indicates the responding 
party---the store. \emph{status} 
indicates the outcome of the execution, since we are not accounting for 
exceptional outcomes, the \emph{status} is \emph{success}.

\subsection{On-Blockchain Deployment}
\noindent The technology that we use in the integration is shown in Fig.~\ref{fig:cccweb3jethereum}.
We have split the contract example into two parts: \emph{dataseller.drl} and
\emph{collectPay.sol}.

\textbf{dataseller.drl} corresponds to the $SC_c$ and is encoded in drools. 
We deploy it on a Mac computer (regarded as a TTP node)  as 
explained in Section~\ref{ccc}.
On the Mac we also deploy an ethereum client connected to the Rinkeby 
Ethereum network (see Fig.~\ref{fig:cccweb3jethereum}).

\begin{figure}[!t]
	\centering
    \includegraphics[width=0.98\columnwidth]{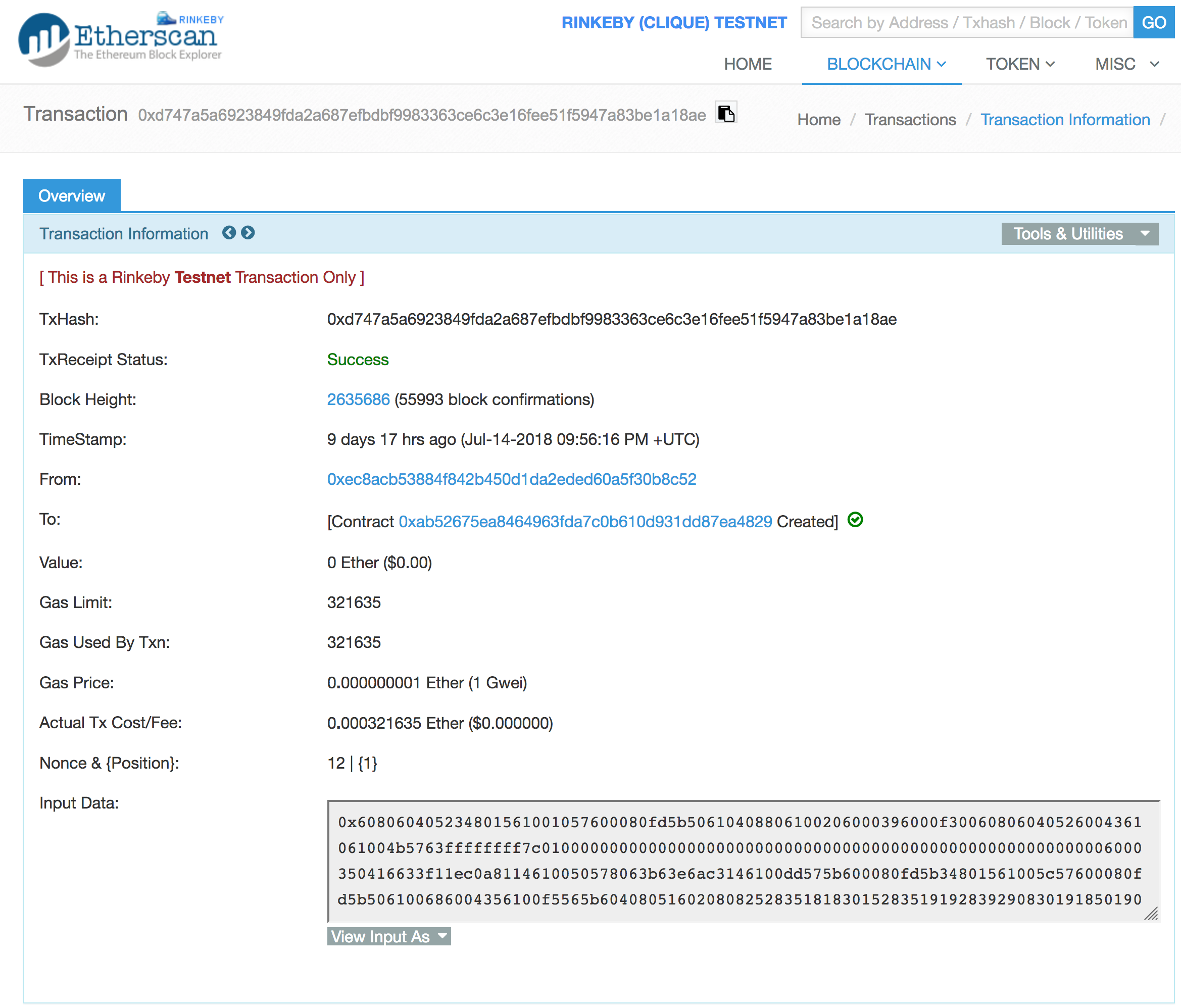}
	\caption{Transaction that deployed the collectPayment.sol contract.}
	\label{fig:contract-deployment-transaction}
\end{figure}

\textbf{collectPayment.sol} corresponds to the $SC_d$ and is encoded in Solidity 
language~\cite{Solidity2017}. 
There are several alternatives such as the \texttt{web3j} library
to deploy the \texttt{collectPayment.sol} contract.
For simplicity, we opted for metamask~\cite{MetaMaskInstall}: a plugin that allows
developers to perform operations against 
Ethereum applications (including contract deployment) from
their browsers, without deploying
a full geth Ethereum node. We deployed metamask on Firefox and, before instantiating
the CCC, we executed the transaction shown in 
Fig.~\ref{fig:contract-deployment-transaction} to deploy the
\texttt{collectPayment.sol} contract on the Rinkeby test network~\cite{CollectPayTx}. We used Ether tokens to pay for gas.
 

\subsection{Smart contracts code}

\noindent The following code contains two rules extracted from the 
\emph{dataseller.drl} contract.

{\scriptsize

\begin{verbatim}
#dataseller.drl contract in drools

rule "Payment Received"
# Grants buyer the right to get a voucher when
# the buyer's obligation to pay is fulfilled.
   when
     $e: Event(type=="PAY", originator=="buyer", 
     responder=="store", status=="success")
     eval(ropBuyer.matchesObligations(payment))
   then // Remove buyer's oblig to pay or cancel
     ropBuyer.removeObligation(payment, seller);
     ropBuyer.removeRight(cancelation, seller);	
     bcEvent.submitPayment();//forward pay to ether contr 
     ropBuyer.addRight(voucher,seller,0,0,120); //5 days
     CCCLogger.logTrace("* Payment result received - 
     add right to GetVoucher ");
     CCCLogger.logTrace("* Payment rule triggered");
     responder.setContractCompliant(true);
end

rule "Get Voucher"
# Grants a voucher to the buyer if the buyer has the right
# ('cos it has fulfilled his payment oblig) to get it.
# It removes the buyer's right to get a voucher after given 
# it to him or 5 days expiry.
   when
     $e: Event(type=="GETVOU", originator=="buyer", 
     responder=="store", status=="success")
     eval(ropBuyer.matchesRights(voucher))
   then
     ropBuyer.removeRight(voucher, seller);
     bcEvent.getVoucher();
     CCCLogger.logTrace("* Get Voucher rule triggered");
     responder.setContractCompliant(true);
end
\end{verbatim}
}

\noindent The following code is the \emph{collectPayment.sol} contract.

{\scriptsize
\begin{verbatim}
///collectPayment.sol contract in Solidity
pragma solidity ^0.4.4;
contract collectPayment{
...    
     
function submitPayment(uint pay) public constant 
returns (string) { 
/// func to submit payment. Returns:
/// "Payment received " + pay converted into str
 var s=uint2str(pay);
 var new_str=s.toSlice().concat("Received".toSlice());
 return new_str;
} 

function getReceipt(uint trasactionNum) public constant 
returns (string) { 
/// func to get a receipt of a given Tx. 
/// returns: "Receipt 4 Tx " + transactionNum 
/// converted into str
 var s=uint2str(trasactionNum);
 var new_str="Receipt 4 Tx".toSlice().concat(s.toSlice());
 return new_str;
 }
}  
\end{verbatim}
}

\noindent Since our focus here is to demonstrate the hybrid architecture, the
\emph{collectPayment.sol} contract is simple: it only receives string messages 
(not Ether tokens or actual Ethereum currency) 
from the \emph{dataseller.drl} contract and replies with another string message.
 
The \emph{client} corresponds to the \emph{client node} of Fig.~\ref{fig:conceptualarchitecture} 
 and acts as a web client to the CCC. We use it to test the implementation of the
 contract example implemented by the combination of  \emph{dataseller.drl} 
 and \emph{collectPayment.sol}. In this order, we provide the client with all 
 the execution sequences encoded in the contract example and previously
 stored in the \emph{dataseller.pmlExecSeq-xml} folder (see Fig.~\ref{fig:folderwithexecsequences}.

\begin{figure}[!t]
	\centering
	\includegraphics[width=0.95\columnwidth]{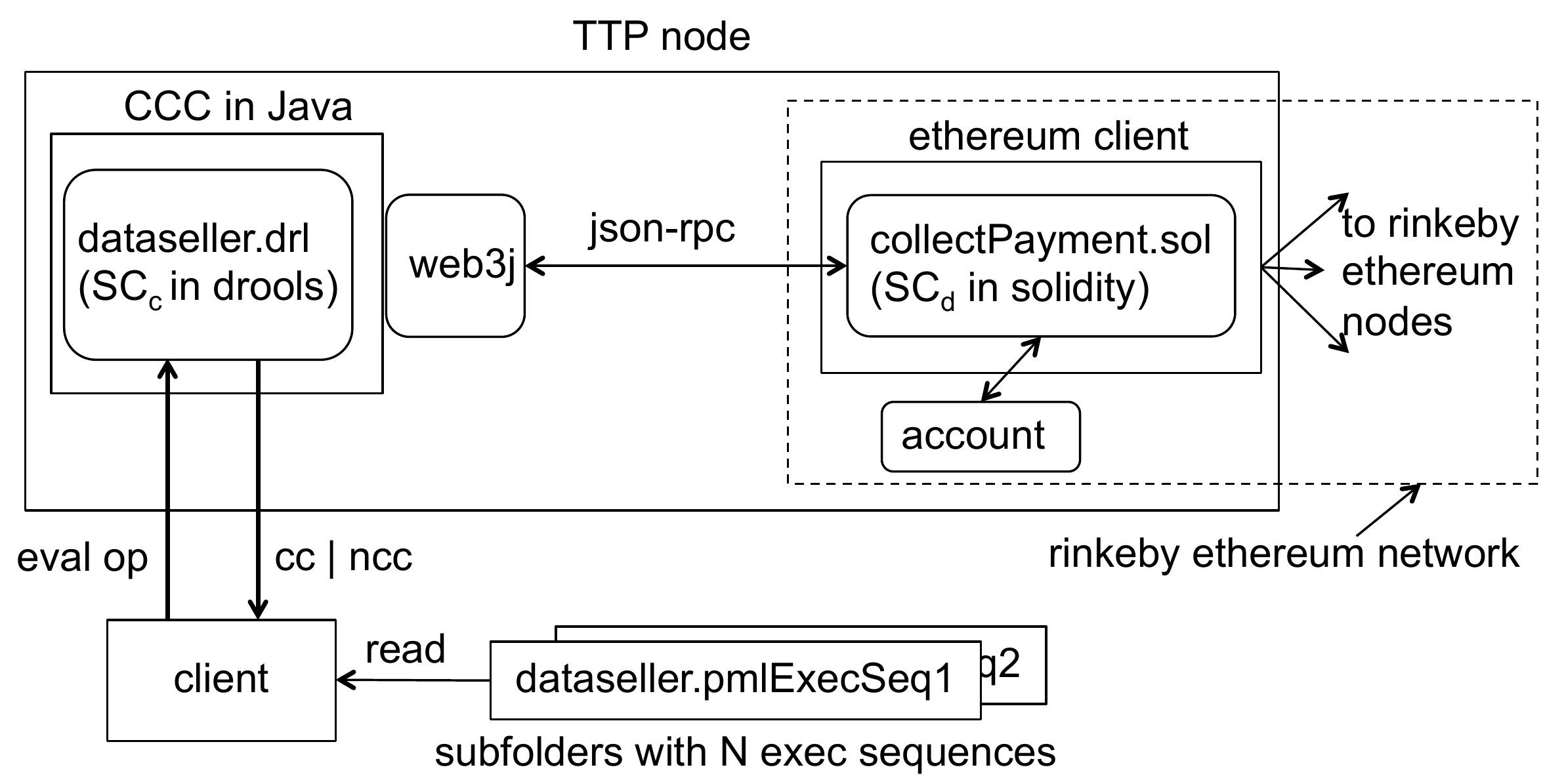}
	\caption{Hybrid architecture for smart contract: technology view.}
	\label{fig:cccweb3jethereum}
\end{figure}

As shown in Fig.~\ref{fig:cccweb3jethereum}, the CCC relies on the \texttt{web3j} library~\cite{web3jio}
to communicate with the Ethereum client.
Among other services, \texttt{web3j} includes a command line application that mechanically generates wrapper code
from a smart contract specified in Solidity and compiled using the solc compiler. The CCC (a Java application) can use the generated wrapper code 
to communicate with the \texttt{collectPayment.sol} contract,
through the json--rpc API provided by ethereum. In addition, the \texttt{web3j} library provides an API for the CCC to unlock an Ethereum client account by providing the path to the keystore file and the password.

In our implementation, the communication facilities provided by
\texttt{web3j} are used by the \texttt{bcEvent.submitPayment()} 
method of the \texttt{dataseller.drl} contract to forward
the \texttt{Pay} operation to the \texttt{collectPayment.sol} 
contract. Intuitively, the statement calls the \texttt{submitPayment}
function of the \texttt{collectPayment.sol} contract. The aim
of this example is to demonstrate how the \texttt{dataseller.drl} and \texttt{collectPayment.sol} 
contracts can communicate with each other. Another 
example of communication is \texttt{bcEvent.getVoucher()} of 
the \texttt{Get Voucher} rule. As it is, this statement calls
the \texttt{getReceipt} function of the \texttt{collectPayment.sol}
contract to receive a string. In production, it could be replaced by a 
function that returns actual Ethers representing the voucher for 
the buyer, or by any other function.

\subsection{Determination of contract compliance}
\label{determination-cc-ncc}

\noindent Let us examine the procedures followed by the \emph{dataseller.drl} and 
\emph{collectPay.sol} contracts to process the operations included
in the contract example. We start with execution sequences that
do not include the \texttt{Pay} operation.

\begin{enumerate}
 \item We assume that the set of the $N$ execution sequences to test the 
       architecture are already available from a local folder
       as explained above. 
  \item We load the CCC with the \emph{dataseller.drl}
        contract and instantiate it to listen for incoming events.
  \item We instantiate the client. In response, it proceeds to
        read the ${dataseller.pmlExeSeq}_1$ folder to extract
        its execution sequence: \texttt{BuyReq, Rej}.
        Next the client sends the \texttt{BuyReq} event to the
        \emph{dataseller.drl}
        contract formatted as a RESTful message.
   \item The \texttt{BuyReq} event triggers a rule 
         of the \emph{dataseller.drl} contract that determines if the
         corresponding \texttt{BuyReq} operation was contract compliant 
         or non contract compliant. The \emph{dataseller.drl}
         contract sends its verdict back to the client.
    \item The above procedure is repeated with the next event (\texttt{Rej})
          of the execution sequence.        
    \item When the client sends the last event of the execution sequence, it
          proceeds to the ${dataseller.pmlExeSeq}_2$, followed
          by the ${dataseller.pmlExeSeq}_3$ and so on, till 
          ${dataseller.pmlExeSeq}_N$. Since all the sequences are legal, the
          \emph{dataseller.drl} contract will declare each event of each sequence
          to be contract compliant.
          
          However, the procedure changes
          when the \emph{dataseller.drl} is presented with an
          event that is meant to be processed by the ethereum \emph{collectPay.sol} 
          contract, such as the \emph{Pay} operation in the
          contract example. Let us discuss this situation separately.
    \end{enumerate}
 
\noindent The execution sequence \texttt{seq: BuyReq, Conf, Pay, GetVou} which
 includes the \texttt{Pay} operation is more problematic because
 it involves the \texttt{collectPayment.sol} contract.
 \texttt{BuyReq} and  \texttt{Conf} are processed by the
 client and \emph{dataseller.drl} contract as above. However,
  when the \emph{dataseller.drl} receives the \texttt{Pay}
  event, the rule  \texttt{Payment Received} (see the
  \emph{dataseller.drl} code) does not evaluate it
  immediately for contract compliance but performs the following
  actions:
          
 \begin{enumerate}     
  \item It creates a blockchain event object. 
  \item It uses the wrapper code (provided by the \texttt{web3j} library)
        to call  the
        \texttt{submitPayment} function of the \texttt{collectPayment.sol}
         contract by means of a json--rpc message. Basically,
         the message forwards the \texttt{Pay} operation from the
         \texttt{dataseller.drl} contract to the
         \texttt{collectPayment.sol} contract.       
          
   \item The result of the call to the \texttt{submitPayment} function
         is not necessarily notified immediately to the \emph{dataseller.drl}
        contract. Consequently, two situations can develop:
         \begin{itemize}
          \item \textbf{a) Pay confirmation precedes GetVou:} The \texttt{dataseller.drl}
                 contract receives pay confirmation and grants the buyer the right to
                 get a voucher. Consequently, when the \texttt{dataseller.drl} eventually
                 receives the \texttt{GetVou} event from the buyer, the operation
                 is declared contract compliant and the voucher is granted.  
                                
          \item \textbf{b) GetVou precedes pay confirmation:} This situation might
                 happen if we assume that the pay confirmation might take
                 arbitrarily long. Because of this, the \texttt{dataseller.drl} 
                 contract receives the \texttt{GetVou} event from the buyer
                 before receiving pay conformation from the \texttt{collectPayment.sol}
                 contract. Consequently, the \texttt{dataseller.drl} contract
                 declares \texttt{GetVou} non--contract compliance--- as far as
                 the \texttt{dataseller.drl} contract is
                 aware of, the buyer does not have the right to get a voucher.  
         \end{itemize}

    
\end{enumerate}



 The materialization of situation \texttt{a)} is shown in the
 outputs produced by the client when it presents
 the \texttt{BuyReq, Conf, Pay, GetVou} sequence to the
 \texttt{dataseller.drl} contract. The text has been
 slightly edited for readability.
 As shown by the \texttt{true} output of 
the third last line, in this execution the 
\texttt{dataseller.drl} contract declares the 
\texttt{GetVou} operation contract compliant.

{\scriptsize
\begin{verbatim}
/* a) In this run of the execution sequence
 *  BuyReq, Conf, Pay, GetVou the dataseller.drl contract 
 * declares the GetVou operation contract compliant: true
 */

 -------- Begin Request to CCC service ----------
BusinessEvent{originator='buyer', responder='store', 
              type='BuyReq', status='success'}
-------- End Request to CCC service ----------

-------- Begin Response from CCC service ----------
 <result>
    <contractCompliant>true</contractCompliant>
</result>
-------- End Response from CCC service ----------


-------- Begin Request to CCC service ----------
BusinessEvent{originator='store', responder='buyer',
              type='Conf', status='success'}
-------- End Request to CCC service ----------

-------- Begin Response from CCC service ----------
 <result>
    <contractCompliant>true</contractCompliant>
</result>
-------- End Response from CCC service ----------


 -------- Begin Request to CCC service ----------

BusinessEvent{originator='buyer', responder='store', 
              type='Pay', status='success'}
-------- End Request to CCC service ----------

-------- Begin Response from CCC service ----------
 <result>
    <contractCompliant>true</contractCompliant>
</result>
-------- End Response from CCC service ----------


 --------- Begin Request to CCC service ----------
BusinessEvent{originator='buyer', responder='store', 
              type='GetVou', status='success'}
-------- End Request to CCC service ----------

-------- Begin Response from CCC service ----------
 <result>
    <contractCompliant>true</contractCompliant>
</result>
-------- End Response from CCC service ----------
\end{verbatim}
}

 The materialization of situation \texttt{b)} is shown in the
 outputs produced by the client when it presents
 the \texttt{BuyReq, Conf, Pay, GetVou} sequence to the
 \texttt{dataseller.drl} contract. 
As shown by the \texttt{false} output of 
the third last line, in this execution the 
\texttt{dataseller.drl} contract declares the 
\texttt{GetVou} operation non contract compliant.
  

{\scriptsize
\begin{verbatim}
 /* b) In this run of the execution sequence
 *  BuyReq, Conf, Pay, GetVou the dataseller.drl contract 
 * declares the GetVou operation non contract compliant: 
 * false
 */

 -------- Begin Request to CCC service ----------
BusinessEvent{originator='buyer', responder='store', 
              type='BuyReq', status='success'}
-------- End Request to CCC service ----------

-------- Begin Response from CCC service ----------
 <result>
    <contractCompliant>true</contractCompliant>
</result>
-------- End Response from CCC service ----------


-------- Begin Request to CCC service ----------
BusinessEvent{originator='store', responder='buyer',
              type='Conf', status='success'}
-------- End Request to CCC service ----------

-------- Begin Response from CCC service ----------
 <result>
    <contractCompliant>true</contractCompliant>
</result>
-------- End Response from CCC service ----------


 -------- Begin Request to CCC service ----------

BusinessEvent{originator='buyer', responder='store', 
              type='Pay', status='success'}
-------- End Request to CCC service ----------

-------- Begin Response from CCC service ----------
 <result>
    <contractCompliant>true</contractCompliant>
</result>
-------- End Response from CCC service ----------


 --------- Begin Request to CCC service ----------
BusinessEvent{originator='buyer', responder='store', 
              type='GetVou', status='success'}
-------- End Request to CCC service ----------

-------- Begin Response from CCC service ----------
 <result>
    <contractCompliant>false</contractCompliant>
</result>
-------- End Response from CCC service ----------
\end{verbatim}
}

\noindent We stress that the problematic situation emerges from
a legal sequence. The potential existence of illegal sequences 
such as those that include \texttt{GetVou} not preceded by
\texttt{Pay} can be uncovered by model checking (for
example, with the \emph{contraval} 
tool) and excluded from the model by the developer. But model checking is not enough. The
error that we are analyzing materialises at run time because
of the interaction (about pay confirmation) between the  
\texttt{dataseller.drl} and \texttt{collectPayment.sol} contracts.
In this work, we uncover it at testing time.

One can also argue that there are simple mechanisms to prevent the
occurrence of the problematic situation (for example, queue the
\texttt{GetVou} event) and to resolve it (for example
the buyer retries the \texttt{GetVou} operation until it is
eventually declared contract compliant by the \texttt{dataseller.drl}
contract. These are valid solutions to the problem; however,
our main observation is that this is only an example of a large
class of situations that might impact hybrid contracts unless 
adequate measures are taken to uncover them at design and testing
time. Such measures need not rely on human labour; static analysis and formal methods should work.

 \subsection{Code and repeatability of experiments}
\noindent The code used in the implementation of Fig.~\ref{fig:cccweb3jethereum} is
 available from the \emph{conch} Git repository. The \texttt{epromela} model and
 the ancillary code used to generate the execution sequences are
 available from the \emph{contraval} Git repository. Interested in readers
 should be able to download both and replicate the experiments
 discussed in this paper.

\section{Executions with abnormal completions}
\label{exceptionaloutcomes}
\noindent For the sake of simplicity, the discussion in 
Section~\ref{modelcheckingtesting} about the contract example
assumes that each contractual operation always succeeds. This is the desirable outcome; however, in practice an operation might fail for business or technical reasons. 
This problem is discussed in~\cite{MolinaTSC2011}.
  
\begin{figure}[!t]
	\centering
	\includegraphics[width=0.75\columnwidth]{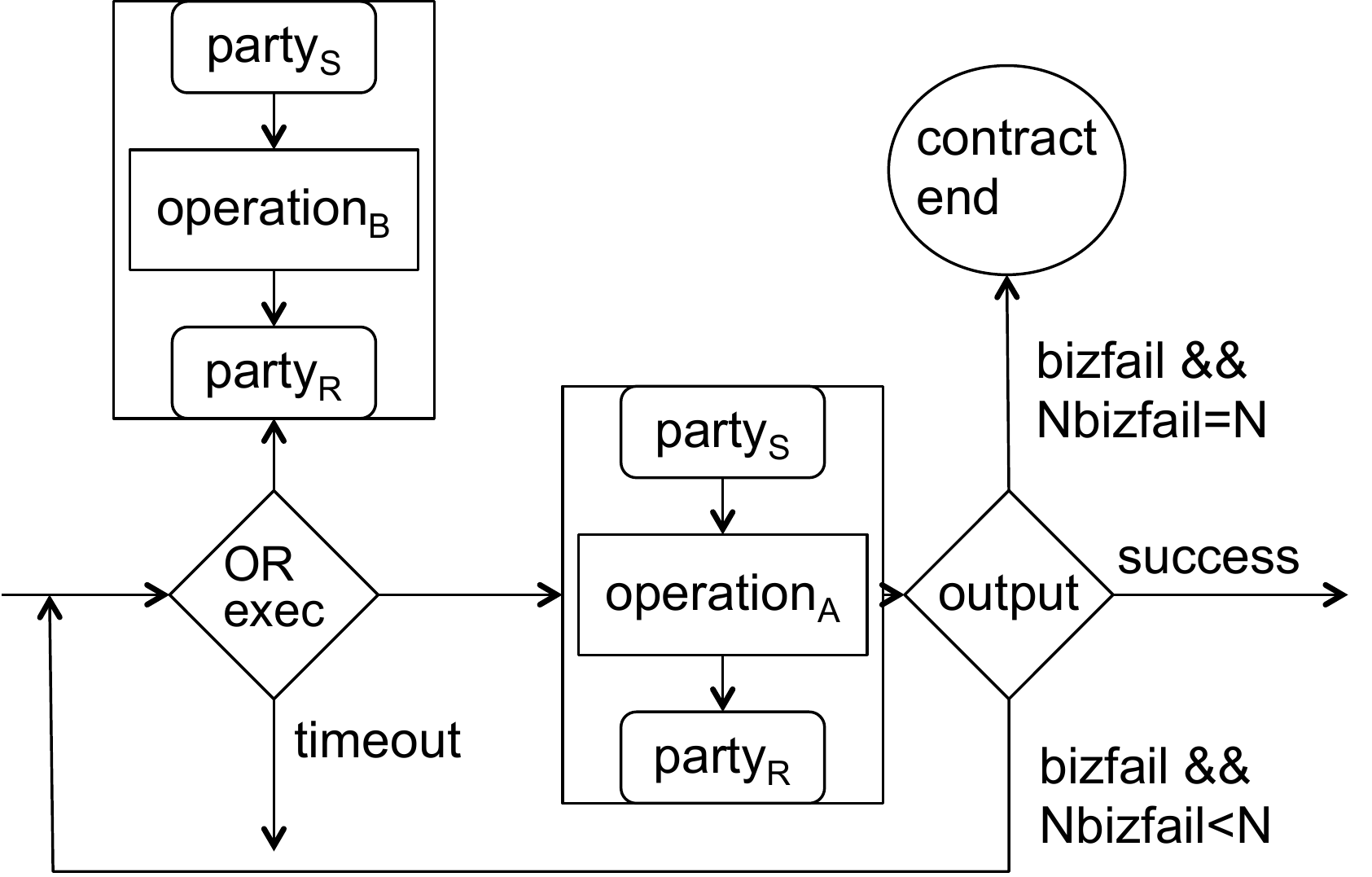}
	\caption{Execution model of contractual operations.}
	\label{fig:operationexecutionmodel}
\end{figure}

  To account for potential exceptional execution outcomes of
  contractual operations, we use the execution model shown
  in Fig.~\ref{fig:operationexecutionmodel}.

  The \emph{OR exec} indicates that there is either a right
  or obligation to execute either $operation_A$ or $operation_B$ before a deadline $timeout$). The timeout arrow leads to the execution of another operation or to the end of the contract. In the simplest case,  $operation_B$ is absent.
  
 Imagine that the $operation_A$ is initiated by $party_S$. The output of the execution can be either $success$ or $bizfail$ (business failure). An execution that completes in $bizfail$ is normally retried until it either succeeds or a number of attempts ($N$) is exhausted. In our execution model,
a $bizfail$ outcome that has not exceeded after $N$ attempts leads to 
the \emph{OR exec} where execution of $operation_A$, $operation_B$ or $timeout$ can take place. It is worth clarifying that the initiator of $operation_A$ and 
$operation_B$ is not necessarily the same contracting party.

We argue that realistic smart contracts need to account for
exceptional outcomes and follow execution models similar to the one shown in Fig.~\ref{fig:operationexecutionmodel}. There are 
several alternatives. For example, a different model results from placing the head of the \texttt{bizfail \&\& Nbizfails <N} to the right of the \texttt{OR exec} diamond. We have build an \texttt{epromela} model of the contract example following the execution model of  Fig.~\ref{fig:operationexecutionmodel}. The \emph{contraval} tool helped us to reveal that the model encodes 246 execution sequences. We have extracted some of them to illustrate our arguments.

{\scriptsize
\begin{verbatim}
// Execution sequences encoded in Fig 1 under the execution
// model of Fig. 4.
// (S)=  execution completed in success
// (BF)= execution completed in business failure
// RejConfTo=Rej or Conf timeout: store failed to execute
//           either rej or Conf by the 3day deadline.
// PayCancTo=Pay or Canc timeout: buyer failed to execute
//           either Pay or Canc by the 7day deadline.
 seq1:   {BuyReq, Rej}
 seq130: {BuyReq(S), Rej(BF), Conf(BF), Conf(S), Pay(BF), 
         Canc(BF), Pay(S), GetVou(S)}
 seq150: {BuyReq(S), Rej(BF), Conf(BF), Conf(S), Pay(BF), 
         Canc(BF), PayCancTO}   
\end{verbatim}
}

Some sequences, like Seq1, are visible to the naked eye. Some can 
be found only by software verification tools. Within this category fall
sequences (for example, \texttt{seq130} and \texttt{seq150}) that include business failures, retries and timeouts.  
\texttt{Seq130} allows the buyer to get his vouchers in spite of the failure of some operations.
However, in \texttt{seq150} the buyer fails to get his voucher due to the occurrence of the \texttt{PayCancTO} timeout. All these sequences are available from 
the \emph{examples/datasellercontract\_TOandBizFailures} folder of the Git \emph{Contraval} repository~\cite{contraval}. Instructions and code to repeat the experiment to generate them are also provided.


\section{Future research directions}
\label{futureresearch}
\noindent We are only starting the exploration of  hybrid 
implementations of smart contracts---our research is at proof of concept stage. To consolidate our ideas, we are planning to conduct a performance evaluation of some QoS requirements to demonstrate that the hybrid approach can meet them, and equally importantly, to demonstrate in which situations the hybrid approach is better than pure off- and on-blockchain platforms. 
For example, we are planning to evaluate the throughput of the channel that communicates the off- and on-blockchain components. In pursuit 
of this aim, we are planning a more demanding
contract example that includes several on-blockchain operations
in addition to the single \texttt{Pay} operation of the current example. Another pending challenge is
the exploration of different interaction models between the 
off- and on-blockchain components. For example, the \texttt{Pay} operation can be sent by the buyer directly to the Ethereum smart contract instead of sending it indirectly through the off-blockchain smart contract. There are also different deployment alternatives for the off-blockchain smart contracts~\cite{MolinaSOCA2011}. For instance, it can be deployed within the buyer's or store's premises instead 
of on a trusted third party.

Our testing does not account for potential failures of the execution
of contractual operations. For instance, it assumes that the pay 
operation always succeeds and
ignores the possibility of technical failures (for example,
the Ethereum network is unreachable) and business failures 
(for example, delivery address not found). We are planning to
explore the behaviour of the hybrid architectures under the
exposure of the execution sequences shown in 
Section~\ref{exceptionaloutcomes}.

Another issue that deserves additional research is the
analysis of the logics implicit in the English text of the
contract as the logics impacts the implementation
complexity and completness of its smart contract equivalent.
The issue is that there several ways of phrasing contractual 
clauses with subtle implications. For instance,
prohibitions can be expressed as obligations. 
Also, the inclusion of a timeout default converts and softens an obligation 
to respond, to a permission to respond.
Finally, in the contract example for 
instance, the buyer's right to obtain a voucher from the 
store could be strengthened to an 
store's obligation to deliver the voucher
if the buyer claims it.
For the sake of readability the contract example is written 
in what is known as
\emph{denormalised} form which correspond to the popular
intuitions about contract deontics~\cite{TomHvitved2012}.
These are issues that we are currently exploring within
the context of our work on programmatic contract drafting.

Programmatic contract drafting is another open research area. In this
regard, we are currently exploring the notion of the Ricardian Contract,~\cite{Grigg}
where systems build and fill templates 
both for formal-language contracts intended for digital execution 
(whether on- or off-blockchain), and for natural-language, human-readable versions
of the contracts. These contracts in natural languages (like the
contract example of this paper) describe the 
same operational core but are 
intended to be signed on paper and legally binding. The natural language contracts also
handle exceptions that cannot be handled
within the on-blockchain contracts; for example, scenarios involving security holes in the on-blockchain contracts, or forks of the blockchain platform itself.
Natural language generation systems offer the potential for 
efficient production and filling of such dual contract templates. 
Together, formal verification of programmatic smart contracts and natural language generation of human-readable contracts promise to create useful synergies: one product of these ideas is a natural language contract which has been mathematically proven to be bug-free.

\section{Related work}
\label{relatedwork}
\noindent Research on smart contracts was pioneered by Minsky in the
mid 80s\cite{Minsky1985} and followed by
Marshall~\cite{Lindsay1993}. Though some of the contract tools
exhibit some decentralised features~\cite{Minsky2010}, those
systems took mainly centralised approaches. Within this category
falls~\cite{GovernatoriEDOC2006} and
\cite{perringodart}. To the same category belongs the model for 
enforcing contractual agreements 
suggested by 
IBM~\cite{ludwig2003soa} and the Heimdahl 
engine~\cite{PedroGama2006} aimed at monitoring state obligations (for example, 
\emph{the store is obliged to maintain the data repository accessible on business days}).
Directly related to our work is the Contract Compliant Checker
reported in~\cite{MolinaTSC2011}~\cite{Solaiman20162} which also took a centralised
approach to gain simplicity at the expense of all the drawbacks that TTPs inevitably introduce.

The publication
of the Bitcoin white paper~\cite{Satoshi2008} in 2008 motivated the development
of several platforms for supporting the implementation of decentralised
smart contracts. Platforms in ~\cite{AndreasAntonopoulos2017}, \cite{Ethereum2017}
and \cite{HyperledgerHome} are some of the
most representative. A good summary of the features offered
by these and other platforms can be found in~\cite{Bartoletti2017}. 
Though they differ in language expressive power, fees and other
features they are convenient for implementing decentralised smart contracts. 

An early example of a permissioned distributed ledger that is similar in functionality to the Hyperledger \cite{HyperledgerHome} of current blockchains is 
\emph{Business to Business Objects (B2Bobjs)}~\cite{NickCook2002}.
\emph{B2Bobjs} is a component middleware implemented at Newcastle University in the early 2000s and used for the enforcement 
of decentralised contracts\cite{Santosh2005}. As such,
it offers consensus services (based on voting initiated by a proposer of a state change) and storage for recording non-repudiable and indelible records of the operations executed by the contracting parties. \emph{B2Bobjs} is
permissioned (as opposed to public) in the sense that only authenticated 
parties are granted access to the object.

  The logical correctness of smart contracts is discussed 
 in several papers~\cite{Christoph2017,Ilya2018,Karthikeyan2018,Anastasia2018}.
 In~\cite{Luciano2016} the author use Petri Nets for validating the correctness of business process expressed in BPMN notation and executed in Ethereum. They mechanically convert BPMN notation into Petri Nets, verify soundness and safeness properties, optimise the resulting Petri Net and convert it mechanically into Solidity. 
 Formal systems for reasoning about the evolution of
 contract executions have also 
 been suggested. Examples of questions of interest are to determine
 the current obligation or state of a party at time $t$ and predicting
 weather a given contract will complete by time $t$. To address
 these issues,  the authors in~\cite{MariaCambronero2017} suggest the
 use of timed calculus to reason about deontic modalities and conditions 
 with time constraints.  A system for programmatic analysis of contracts written in natural languages
 (normative texts) to extract contractual commitments (what parties are and are 
 not expected to do) are discussed in~\cite{Camilleri2017}.

The hybrid approach that we dicusss in this paper
addresses problems that neither the centralised
nor decentralised approach can address separately. It
was inspired by the arguments presented in~\cite{CarlosIoannisTurin2018}, though the original idea emerged by the off-blockchain payment 
channel discussed in~\cite{Joseph2016,AndreasAntonopoulos2017}.
The concept of logic-based smart contracts discussed in~\cite{Florian2016}
has some similarities with our hybrid approach. They suggest
the use of logic-based languages in the implementation of
 smart contracts capable of performing on-blockchain and
 off-blockchain inference. The difficulty with this approach is
 lack of support of logic-based languages in current blockchain technologies.
 In our work, we rely on the native languages offered
 by the blockchain platforms --- here, Ethereum's Solidity.
 On- and off-blockchain enforcement of contractual operations
 is also discussed in~\cite{Xiwei2016}. Though an architecture is
 presented, no technical details about its implementation or
 functionality are given. 
 Another conceptual design directly related to our work is private contracts executed in the Enigma~\cite{Guy2015} architecture. 
As in our work, a private contract is a conventional
business contract with operations separated into on- and off-blockchain categories. Similarly to our hybrid
design, they use a blockchain platform (namely Ethereum) to execute on-blockchain  operations. Unlike in our work, instead of using a TTP to execute off-blockchain
 operations, they use a set of distrusting Enigma nodes 
 running a Secure Multi-party Computation (SPC) protocol~\cite{Yao1982,Marcin2014} that guarantees privacy.
In that collaborative architecture, the blockchain is in charge. It is responsible for guaranteeing that the contractual operations are honoured and for delegating tasks to the Enigma nodes as 
needed. The integration of the SPC protocol ensures that
the smart contract running in the Ethereum blockchain never 
accesses raw data that might compromise privacy. Unlike our TTP,
the Enigma nodes charge computation and storage fees, as
Ethereum and Bitcoin do. The cost that the Enigma architecture pays for privacy protection is complexity.

 The idea of interconnecting smart contracts to enable
 them to collaborate with each other is also discussed
 in~\cite{Hardjono2018}. These authors draw a similarity
 between blockchains and the Internet. They speculate that in
 the future, we will have islands of blockchain systems interconnected by gateways.


\section{Concluding remarks}
\label{conclusions}
The aim of this paper has been to argue that there are good reasons to consider hybrid architectures composed of off- and on-blockchain components as alternatives for the implementation of smart contracts with strict QoS requirements. As a proof of concept, we have demonstrated that hybrid architectures are implementable as long as the off-blockchain component provides standard APIs to communicate with the standard APIs that current blockchains offer.

We have presented the approach as a pragmatic solution to the current problems that afflict off- and on-blockchain platforms. However,
we believe that these ideas will become useful in the development 
of smart contract applications in the near future. We envision cross-smart contract applications that will involve several smart contracts running on  independent platforms. The
architecture that we have implemented is in line with this
vision. Though the current implementation includes only two components, it can be generalised to include an
arbitrary number of off-blockchain and on-blockchain components. This generalisation should be implementable provided that the components offer interfaces (gateways) to interact with each other and the developer devises mechanisms for coordinating their collaboration.  

We have argued that the implementation of sound smart contracts is not trivial and that the inclusion of off- and on-blockchain components makes the task even harder. To ease the task, we advise the use of software tools to mechanise the verification and testing 
of smart contracts.



\section*{Acknowledgements}
Carlos Molina-Jimenez is currently collaborating with the HAT Community 
Foundation under the support of Grant RG90413 NRAG/536. 
Ioannis Sfyrakis was partly supported by the EU Horizon 2020 project 
PrismaCloud (https://prismacloud.eu) under GA No. 644962.
Meng Weng Wong is a 2017--2018 Fellow at Stanford University's CodeX Center for Legal Informatics, and previously a 2016--2017 Fellow at Harvard University's Berkman Klein Center for Internet and Society, and a 2016 Fellow at Ca'Foscari University of Venice.

\bibliographystyle{IEEEtran}
\bibliography{references.bib}
\end{document}